\documentclass[conference]{IEEEtran}
\IEEEoverridecommandlockouts
\usepackage{cite}
\usepackage{amsmath,amssymb,amsfonts}
\usepackage{algorithmic}
\usepackage{graphicx}
\usepackage{textcomp}
\usepackage{xcolor}
\usepackage{url}
\usepackage{booktabs}
\usepackage{multirow}
\usepackage{chngpage}
\usepackage{tabularx}

\usepackage{multirow}
\usepackage[normalem]{ulem}
\useunder{\uline}{\ul}{}
\def\BibTeX{{\rm B\kern-.05em{\sc i\kern-.025em b}\kern-.08em
    T\kern-.1667em\lower.7ex\hbox{E}\kern-.125emX}}
    
\usepackage{,amsmath,graphicx,mathtools,}

\begin{document}

\title{Source Separation-based Data Augmentation for Improved Joint Beat and Downbeat Tracking
}

\author{\IEEEauthorblockN{Ching-Yu Chiu}
\IEEEauthorblockA{\textit{Graduate Program of Multimedia Systems and Intelligent Computing} \\
\textit{National Cheng Kung University and Academia Sinica}, Taiwan\\
sunnycyc@citi.sinica.edu.tw}
\and
\IEEEauthorblockN{Joann Ching}
\IEEEauthorblockA{\textit{Research Center for IT Innovation} \\
\textit{Academia Sinica}, Taiwan \\
joann8512@citi.sinica.edu.tw}
\and
\IEEEauthorblockN{Wen-Yi Hsiao}
\IEEEauthorblockA{\textit{Yating Music Team} \\
\textit{Taiwan AI Labs}, Taiwan\\
wayne391@ailabs.tw}
\and
\IEEEauthorblockN{Yu-Hua Chen}
\IEEEauthorblockA{\textit{Yating Music Team} \\
\textit{Taiwan AI Labs}, Taiwan\\
cloudstrife60138@gmail.com}
\and
\IEEEauthorblockN{Alvin Wen-Yu Su}
\IEEEauthorblockA{\textit{Dept. CSIE} \\
\textit{National Cheng Kung University}\\
alvinsu@mail.ncku.edu.tw
}
\and
\IEEEauthorblockN{Yi-Hsuan Yang}
\IEEEauthorblockA{\textit{Research Center for IT Innovation}\\
\textit{Academia Sinica}, Taiwan \\
yang@citi.sinica.edu.tw}
}

\maketitle

\begin{abstract}
Due to advances in deep learning, the performance of automatic beat and downbeat tracking in musical audio signals has seen great improvement in recent years.  In training such deep learning based models, data augmentation has been found an important technique. However, existing data augmentation methods for this task mainly target at balancing the distribution of the training data with respect to their tempo.
In this paper, we investigate another approach for data augmentation, to account for the composition of the training data in terms of the percussive and non-percussive sound sources. 
Specifically, we propose to employ a blind drum separation model to segregate the drum and non-drum sounds from each training audio signal, filtering out training signals that are drumless, and then use the obtained drum and non-drum stems to augment the training data.
We report experiments on four completely unseen test sets, validating the effectiveness of the proposed method, and accordingly the importance of drum sound composition in the training data for beat and downbeat tracking.

\end{abstract}


\section{Introduction}
Beats and downbeats are usually referred to as a sequence of time instants that human would tap their feet to \cite{Bock2014, Bock2016a}. Beats and downbeats define the metrical structure of a musical piece, and enable numerous higher level tasks such as structure segmentation, automated DJ mixing, and score alignment \cite{Zapata2013,Grohganz2014,Bock2016a}. Moreover, they are  essential foundations for machines to understand and analyze music. As humans are able to track musical beats without difficulty in various musical genres regardless of the existence of drum sounds in an audio signal, people expect machines to have similar skills.
Thanks to cumulative efforts in the research community, recent years witness great success of deep learning-based supervised models for \emph{joint beat and downbeat tracking}---i.e., a single model that estimates beats and downbeats at the same time---for music with steady and strong beats, such as pop, rock, and dance music \cite{Bock}. 
We also focus on the joint beat/downbeat tracking task  in this paper.


Existing deep learning-based methods for beat/downbeat tracking are mainly supervised models that require labeled training data, namely audio with annotated labels of beat and downbeat timing. These approaches usually consist of a neural network module and a post-processing dynamic model. The neural network module takes low-level features such as chroma, spectral flux, or first-order derivatives of magnitude spectrograms \cite{Durand2015,Bock2016a,Krebs2016,Fuentes2018} as input, and outputs an activation function that indicates the most likely beats and downbeats time candidates. The post-processing model, be it a dynamic Bayesian network (DBN), hidden Markov model (HMM) or conditional random field (CRF), makes binary predictions from the activation function \cite{Bock2016a,Durand2017,Fuentes2018}. 

Challenges for beat/downbeat tracking can be viewed from the inherent signal characteristics for different type of music, such as strong tempo variation, various tempo distribution, rhythmic variation or syncopation for purposes of creating metrical tension, and the rather blurred note onsets and transitions caused by non-percussive instruments  \cite{syncopation, Hockman2012, Grosche2010, Ellis2014}. From a machine learning perspective, the difficulty may also be related to limitations of model capacity, insufficient labeled data for training, or potential imbalance of training data. For example, we note that most available training datasets for beat/downbeat tracking are pop/rock songs \cite{Bock}. Moreover, while deep learning models perform well, the performance of a deep learning model for beat/downbeat tracking seems to be sensitive to the composition of training data. We found the training models with the same model architecture, comparable training data size, yet different data constitution (e.g., tempo distribution, percussive/non-percussive sound source  percentage) can lead to  significant performance difference on different test sets. 

Although not often mentioned together with beat tracking, music source separation has been another important topic in music information research. Given a monaural input mixture (i.e., an audio recording of multiple instruments sounding together), a source separation model generates separated stems of different sound sources that composed the mixture \cite{Uhlich2017,Rafii2018,Hennequin2019,Stoter2019,Chiu2020}. Noticing that the total duration and metrical structure of the stems are basically the same, we conjecture that it may be useful to apply drum/non-drum  separation as a means of data augmentation, to increase the amount of drum or non-drum data in our training set. It is therefore our goal in this paper to investigate variants of such augmentation data and their usefulness in improving the performance of joint beat and downbeat tracking. To the best of our knowledge, such a study has never been presented before.

As our focus is on data augmentation, for the model architecture, we simply adopt the pipeline of cascading a long short-term memory (LSTM) network with an HMM-based dynamic network proposed by \texttt{Madmom} \cite{Bock2016,Bock2016a}, which represents the state-of-the-art for joint beat and downbeat tracking across various genres.
We evaluate the performance of our models on four completely unseen datasets in our experiments, namely ASAP  \cite{asap-dataset}, Rock    \cite{rockdataset}, HJDB \cite{Hockman2012}, and RWC Royalty-Free \cite{rwcdatabase}. We plan to open source our code at \url{https://github.com/SunnyCYC/aug4beat}.


\section{Background}
\label{sec:format}


As the available labeled data for training supervised deep learning models for beat tracking is limited, research has been done to enhance the generalizability of models, or to compensate for the relatively insufficient properties in the training set. Giorgi \emph{et al.}  \cite{Giorgi}, for example, proposed a deterministic time-warping operation to help their model learn rhythmic patterns independently of tempo. B{\"{o}}ck and Davies \cite{Bock} devised a novel multi-task approach to leverage shared connections in musical structure, and to simultaneously estimate tempo, beat, and downbeat. They also proposed an augmentation method based on changing parameters of the short-time Fourier transform to  expose their model to a wider range of tempi. Zapata and Gomez \cite{Zapata2013} proposed an audio voice suppression technique and a simple low-pass filter to improve beat tracking.

Efforts have been made to tackle beat tracking by treating percussive/non-percussive features separately. Goto \cite{ Goto2001} developed a method to judge if the input audio signal contains drum sounds, and then used different ways to track the beats for music with or without drum sounds. Gkiokas \emph{et al.} \cite{Gkiokas2012} presented an tempo estimation and beat tracking algorithm which utilized source separation to extract features of percussive/harmonic components separately. 
Our work is different from these prior arts in that we use source separation as a means to create augmented data for training a supervised neural network model.

We implement a beat/downbeat tracking model on our own following the LSTM$+$HMM architecture described in the \texttt{Madmom} library \cite{Bock2016,Bock2016a}. The network consists of three fully-connected bidirectional recurrent layers with 25 LSTM units each. After the LSTM layers, a softmax classification layer with three units produces three activation functions (i.e., curves) corresponding to the probability of a frame being `a beat but no a downbeat,' `a downbeat' or `a non-beat' position. The output activation functions are then processed by an HMM to produce the final binary beat/downbeat predictions. 
In other words, the model predicts beat and downbeat jointly.
In this paper, the bar length setting required by HMM is set as three or four beats, following the default setting of \texttt{Madmom}.

\begin{table}
\caption{The datasets employed for model training (top) and testing (bottom) in our experiments; `\#' denotes the number of musical pieces. The last two columns indicate the percentage of the selected drum stems for the dataset. Most of the 10 datasets listed here have the drum stems after source separation, except for RWC Classical \cite{rwcdatabase} and ASAP \cite{asap-dataset}, which has no presence of drum sounds at all.
}
\centering
\begin{tabular}{l|rrrrr}
\toprule
\multirow{2}{*}{\textbf{Dataset}} & \multirow{2}{*}{\#} & Total  & 
\multicolumn{2}{c}{drum presence rate} \\
\cline{4-5}
&  & duration & ABSM & OSFQ\\
\midrule
RWC Classical \cite{rwcdatabase}  & 54  & 5h\,19m  & 0.0\%  & 0.0\%  \\
RWC Jazz Music \cite{rwcdatabase}  & 50  & 3h\,42m  & 29.8\% & 64.1\% \\
RWC Music Genre \cite{rwc_musicgenre} & 100 & 7h\,20m  & 39.3\% & 62.6\% \\
Ballroom \cite{ballroom1, ballroom2} & 685 & 5h\,57m  & 59.1\% & 82.2\% \\
Hainsworth \cite{hainsworth} & 222 & 3h\,19m  & 65.0\% & 75.1\% \\
GTZAN \cite{gtzan1, gtzan2} & 999 & 8h\,20m  & 71.2\% & 79.0\% \\
Carnatic \cite{carnatic}  & 176 & 16h\,38m & 72.8\% & 84.2\% \\
Beatles  \cite{beatles}       & 180 & 8h\,09m   & 75.8\% & 94.0\% \\
RWC Popular \cite{rwcdatabase}   & 100 & 6h\,47m  & 89.5\% & 90.8\% \\
Robbie Williams \cite{robbiewilliams}  & 65  & 4h\,31m  & 93.5\% & 98.5\% \\
\midrule
ASAP  \cite{asap-dataset}  & 520 & 48h\,07m & 0.0\%  & 0.0\%  \\
Rock    \cite{rockdataset}  & 200 & 12h\,53m & 83.2\% & 97.1\% \\
HJDB    \cite{Hockman2012}  & 235 & 3h\,19m  & 99.2\% & 99.2\% \\
RWC Royalty-Free \cite{rwcdatabase} & 15  & 27m    & 99.9\% & 99.9\%\\
\bottomrule
\end{tabular}
\label{tab:dataset_overview}
\end{table}

\section{Proposed Method}
\label{sec:pagestyle}
\subsection{Source Separation-based Data Augmentation}
We propose to employ a source separation model to isolate out the drum sounds and non-drum sounds for the musical pieces in the training data for beat/downbeat tracking. 
In doing so, we employ Spleeter \cite{Hennequin2019}, the current state-of-the-art model for musical source separation that has publicly available model checkpoints. 
Spleeter can generate four separated stems, each stem corresponding to the sounds of the vocals, drums, bass, and others, respectively. We simply sum all but the drum stem as the \emph{non-drum} stem. 
If we call the original pieces in the training set as \emph{mixes}, we would get the same number of drum stems and non-drum stems after source separation. For those pieces that correspond to the same song, we assume that they share the same metrical structure. Accordingly, we can assume that the labels of beat and downbeat timing apply equally well to that corresponding drum stem and non-drum stem.

\subsection{Drum Stem Selection}
\begin{figure}
\centering
\includegraphics[width=0.9\columnwidth]{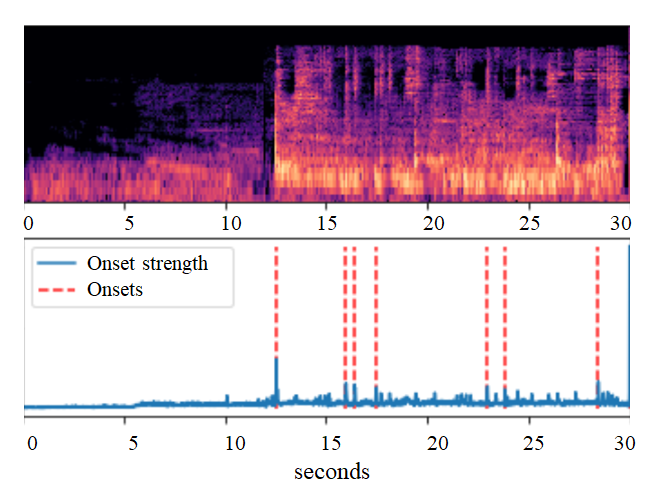}
\caption{An example of a separated drum stem that would pass the ABSM filter as it has  recognizable drum sounds, yet would not pass the OSFQ filter, for it lacks regular beat structure as can be seen from the result of onset detection in the bottom figure. The top figure shows its spectrogram.
}
\label{fig:onset_detection}
\end{figure}

Since some of our training data may not contain any drum sounds originally, the drum stems derived from them would be nearly silent and could be harmful if adopted as training data. We devise two drum selection criteria to tackle this. 
\noindent \textbf{ABSM}: Exclude a drum stem if the mean value of its absolute magnitude in the time domain is less than 0.01  (while the maximum absolute magnitude is 1), an empirically set threshold. 
\noindent \textbf{OSFQ}: Exclude a drum stem if the mean value of its absolute magnitude is less than 0.001 (i.e., looser than ABSM), \emph{or} if it has less than one prominent onset per second, as detected by \texttt{librosa.onset.onset\_detect} \cite{McFee2015}---i.e., when its onset patterns is sparse or irregular. See Figure \ref{fig:onset_detection} for an example of bass drum excluded by OSFQ.

After drum selection, we have in total four types of data: \emph{Mix}, \emph{non-drum}, \emph{only-drum (ABSM)} and \emph{only-drum (OSFQ)}, for all the  pieces in our training set. 
To manipulate the percentage of drum sounds in our training set as data augmentation, we can use different combinations of them. For example, we can use \emph{Mix} with \emph{non-drum}/\emph{only-drum} to reduce/increase the drum sound percentage of our training data, while both usage can result in comparable total amount of training data. Please see Table \ref{tab:dataset_overview} for the datasets employed in training our models.


\section{Experiment}
\label{sec:exp}


We implement models with five data type combinations that are enabled by source separation, as listed in Table \ref{tab:combination_trainin}.
The goal is to see any of these can outperform the baseline case of using the \emph{Mix}  alone for model training, as it represents the standard approach as of now.
For these models, we randomly split the top 10 datasets listed in Table \ref{tab:dataset_overview} into 80\%, 10\%, 10\% for training, validation and test. During validation, the parameters of the post-processing HMM model are optimized. The proposed data augmentation applies to the training sets only. To alleviate the random effect, we repeat the training of all models for five times and report the averaged evaluation results. Due to space limit, we only report the results on the four unseen test sets on the bottom of Table \ref{tab:dataset_overview}. The result on the testing split of the top 10 datasets is consistent with that on the unseen test sets.

We consider the model trained with the original mixes as the \emph{baseline} in our experiments. We call it  ``Madmom-like,'' for it uses the same 3-layer bidirectional LSTM$+$HMM archecture as the RNNDownBeatProcessor of  Madmom  \cite{Bock2016,Bock2016a}. 



\begin{table}
\caption{The data type combinations tested in our experiments, along with the short names of the corresponding models. Each model is trained from scratch using the ticked data types of the training split of the 10 datasets listed in Table \ref{tab:dataset_overview}. }
\centering
\begin{tabular}{l|cccc}
\toprule
\multirow{3}{*}{\textbf{Model}}  & \multicolumn{4}{c}{\textbf{Data}}

\\
& \multirow{2}{*}{mix}   & \multirow{2}{*}{non-drum} & \multicolumn{2}{c}{onlyDrum} \\
\cline{4-5}

& &  &  ABSM  & OSFQ \\
\midrule
Mix (baseline)                 & $\surd$ &   &   &    \\
\hline
Mix$+$no-drum      & $\surd$ &$\surd$ & & \\
Mix$+$only-drum\_ABSM        &$\surd$   &  & $\surd$ &   \\
Mix$+$only-drum\_OSFQ       &$\surd$   &  &   & $\surd$ \\
\hline
3combABSM    & $\surd$ & $\surd$ & $\surd$ &   \\
3combOSFQ     & $\surd$ & $\surd$ &   & $\surd$\\
\bottomrule
\end{tabular}

\label{tab:combination_trainin}
\end{table}


\begin{table*}[]
\caption{The resulting F1 scores (averaged over five runs) in beat and downbeat tracking in four different test sets unseen at training time. Each model is trained from scratch using the ticked data types, shown in Table \ref{tab:combination_trainin}, of the training split of the 10 datasets listed in Table \ref{tab:dataset_overview}. In each column, we highlight the result that outperforms baseline in bold, and the best result with underline. }
\centering
\begin{tabular}{l|cc|cc|cc|cc}
\toprule
\multirow{2}{*}{Model} & \multicolumn{2}{c|}{ASAP} & \multicolumn{2}{c|}{Rock}        & \multicolumn{2}{c|}{HJDB}              & \multicolumn{2}{c}{RWC Royalty-Free}        \\
          & beat F1        & downbeat F1    &beat F1                 & downbeat F1             & beat F1           & downbeat F1       & beat F1           & downbeat F1       \\
\midrule
Madmom \cite{Bock2016,Bock2016a}-like baseline  & {\ul 0.591} & {\ul 0.406} & 0.874 & 0.788          & 0.882                & 0.679                & 0.833          & 0.810                \\
\hline
Mix$+$no-drum & 0.586       & 0.398       & 0.869 & \textbf{0.796} & \textbf{0.900}       & \textbf{0.685}       & \textbf{0.861} & {\ul \textbf{0.853}} \\
Mix$+$only-drum\_ABSM  & 0.589       & 0.402      & {\ul \textbf{0.882}} & {\ul \textbf{0.800}} & \textbf{0.896} & \textbf{0.735} & {\ul \textbf{0.865}} & \textbf{0.852} \\
Mix$+$only-drum\_OSFQ              & 0.590       & 0.401       & 0.874 & \textbf{0.790} & \textbf{0.894}       & {\ul \textbf{0.771}} & \textbf{0.836} & \textbf{0.826}       \\
3combABSM & 0.583       & 0.394       & 0.870 & \textbf{0.794} & \textbf{0.886}       & \textbf{0.758}       & \textbf{0.837} & \textbf{0.812}       \\
3combOSFQ & 0.581       & 0.399       & 0.868 & 0.783          & {\ul \textbf{0.902}} & \textbf{0.757}       & \textbf{0.847} & \textbf{0.843}      \\
\bottomrule
\end{tabular}
\label{tab:results}
\end{table*}

\subsection{Result on the Four Unseen Test Sets}
\label{ssec:subhead}


Table \ref{tab:results} shows the F-measure (F1) results of the proposed methods. The following observations can be made. Except for case of the ASAP test set, which is composed of expressive classical piano performance, the augmented models obtain different levels of improvement in the other three test sets. 
We speculate that this is due to the fact that the dataset characteristics (e.g., local tempo variation, rhythmic pattern, note onset/offset patterns) of ASAP is quite different from most of our training sets. As a result, even though the \emph{non-drum} stems are adopted in \emph{Mix$+$no-drum} to increase the percentage of non-drum audio signals in the training set, the difference between the \emph{non-drum} of the train sets and ASAP still deteriorate the performance. 

For results of HJDB ---a drum-heavy dataset comprising Hardcore, Jungle and Drum\&Bass music excerpts \cite{Hockman2012}, models augmented by \emph{onlyDrum} stems gains clear improvement (e.g., at least $+$5.5\% relative improvement in downbeat), while model augmented by \emph{non-drum} stem gets relatively limited improvement.  Such an observation implies the noticeable influence of trainset sound source composition on beat/downbeat tracking performance on different test sets. We also notice the unexpected downbeat improvement of \emph{Mix$+$no-drum} on RWC Royalty-Free. One reason for this lies within the difference between HJDB and RWC Royalty-Free. While songs in HJDB are dominanted by drum sound all the time, songs in RWC Royalty-Free may be without drum sound for several measures. 


The performance improvement on the Rock dataset is less obvious than that on either HJDB or RWC Royalty-Free. This may be due to the fact that the majority of our training data (as shown in Table \ref{tab:dataset_overview}) is composed of rock/pop music---the baseline model may have learned most of the beat/downbeat related patterns required for tracking Rock music. The downbeat improvements are consistently larger than the beat ones, which may suggest downbeat gains more advantage from the proposed data augmentation.  

In summary, it is clear that the test sets with high drum presence rate can benefit from the extra stems derived from source separation and drum selection. And the exact best combinations/composition of training data depends on the characteristics of target test set. 

\subsection{Case Study: Analysis of Improvement on HJDB }
\label{ssec:subheadcase}
\begin{figure}
\centering
\includegraphics[width=1.0\columnwidth]{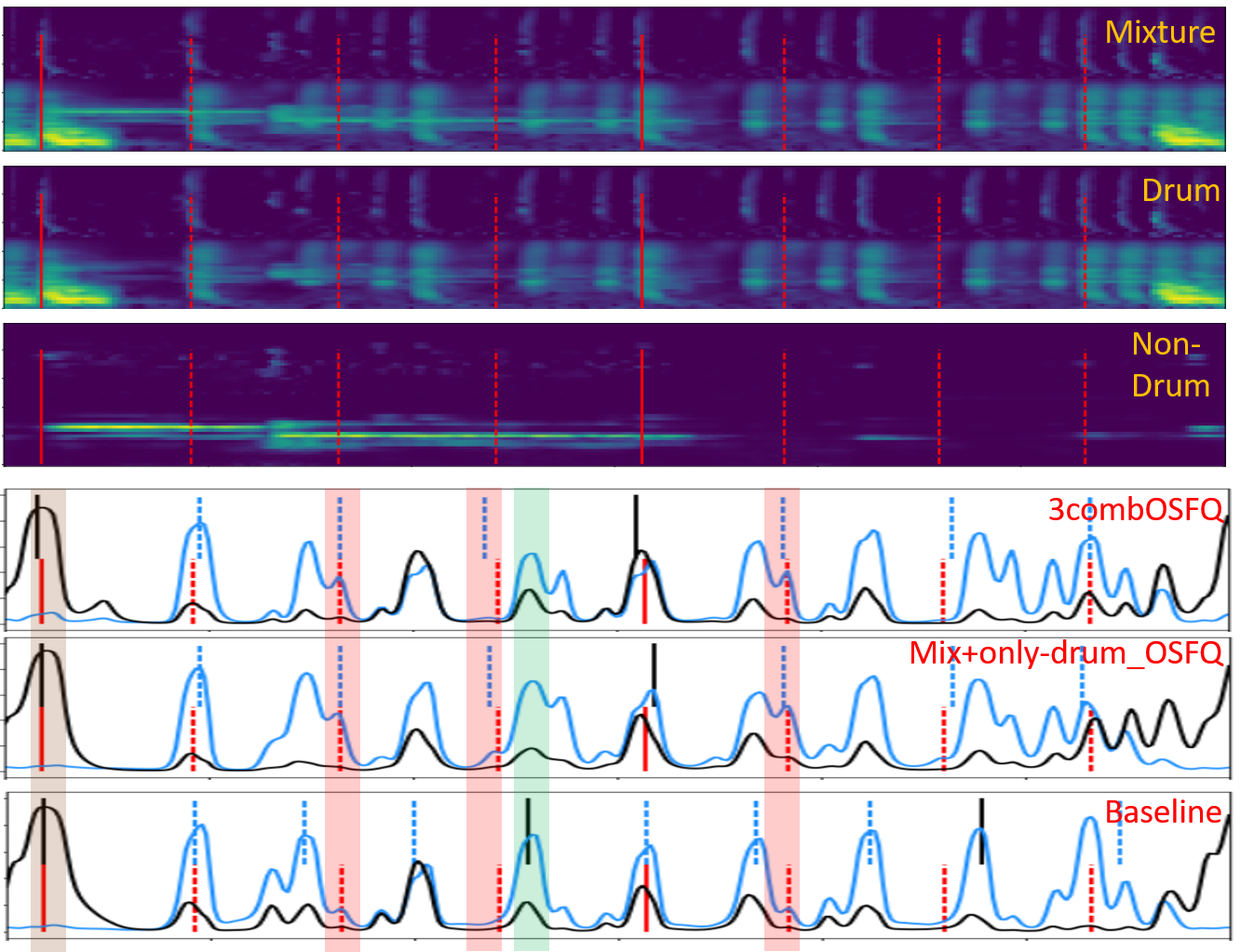}
\caption{Estimation (vertical lines) and activation (curves) of beat (blue) and downbeat (black) on a clip of a song (cut from 17--21 seconds) from HJDB \cite{Hockman2012}, named ``Sweet Vibrations.'' The input features (magnitude spectrogram and its first order derivative) calculated from mixture, drum stem, and non-drum stem are also partially shown for comparison. Red vertical lines indicate the ground-truth annotation of beats (dashed) and downbeats (solid). In this example, the beat/downbeat estimation (finalized by HMM) of the augmented models, \emph{3combOSFQ} and \emph{Mix$+$only-drum\_OSFQ}, are comparable and are closer to ground-truth, while the \emph{baseline} model gradually deviates from the ground-truth annotations. Brown square on the left hand side highlights the correct estimation of downbeats and ensures the latter errors are not propagated from previous unshown part of the song. Red/green squares highlight the critical positions of beats/downbeats that cause the failure of \emph{baseline}. The difficulty of this song mainly comes from the frequent syncopation (i.e., the drum onsets between beats), which could cause false-positive activation peaks between beats. Although all the three models exhibit such false-positive activation peaks, the augmented models somehow learn the idea of syncopation and are able to generate activation peaks at the nearly silent beat positions, enabling the HMM to `correct' the final results. 
}
\label{fig:case}
\end{figure}

To have an idea of the underlying causes for improvements on HJDB, we compare the model activation functions, beat/downbeat estimations/annotations and corresponding input feature, and show one representative example from HJDB in Figure \ref{fig:case}. From the top three feature plots, we can observe the following features of this song: the relative sparseness of non-drum feature reflects the drum sound dominance. The number of beats in a time span of three seconds, and the several drum onsets in between beats indicate this is a fast tempo song with frequent use of syncopation. These are features distinguishing HJDB from the other datasets we employed. From the beat/downbeat estimations (blue/black vertical lines), we can see that the drum stem-augmented models (i.e. \emph{3combOSFQ} and \emph{Mix$+$only-drum\_OSFQ}) lead to predictions that are closer to the ground-truth (red vertical lines) than the baseline. However, such an improvement can not be achieved without the help of HMM. From the several `false-positive' activation peaks between beats produced by the three models, we can see the syncopation clearly `misleads' the three models. As the underlying mechanism of HMM decoding is to find the global optimal path, HMM has the chance to `correct' the local activation error as long as the model also produces peaks high enough at the right beat positions. From the three red squares indicating the critical beat positions, we can see that \emph{Mix$+$only-drum\_OSFQ} is able to generate peak at the right positions, while \emph{3combOSFQ} fails for the second square but benefits from HMM. And the failure of \emph{baseline} to generate clear peaks at the three positions not only prevents it from the help of HMM but also deteriorates HMM's judgement on downbeat (highlighted by green square). Looking back to the drum feature again, it seems reasonable that the augmented models may have better chance than \emph{baseline} to learn the idea of syncopation. While the truth beat/downbeat positions may fall in nearly silent positions (e.g., between drum onsets) in drum stems, such cases are less frequent in the original audio mixture due to the presence of other non-drum sounds. In summary, this example demonstrates the underlying reasons for the performance improvement derived by the proposed methods, and reveals that the HMM plays some role in taking advantage of the improved network activation prediction.

\section{Conclusion}

In this paper, we have presented  a source separation-based data augmentation technique that utilizes the drum/non-drum stems separated out by source separation to increase the presence of drum-only and non-drum data in the training set. 
We have also reported experiments showing that 
the performance of joint beat/downbeat tracking can depend on the sound source composition of the training set, and that the proposed data augmentation method leads to performance improvement across different test sets. For future work, we are interested in building a beat/downbeat tracking model that incorporates source separation as part of the model, and in conducting more experiments on classical music pieces.



\bibliographystyle{IEEE}
\bibliography{eusipco2021}


\end{document}